# On the Analysis and Design of High-Frequency Transformers for Dual and Triple Active Bridge Converters in More Electric Aircraft

Babak Rahrovi, *Student Member, IEEE*, Ramin Tafazzoli Mehrjardi
and Mehrdad Ehsani, *Life Fellow, IEEE, Fellow, SAE*
Department of Electrical and Computer Engineering, Texas A&M University, College Station, Texas, USA
brahrovi@tamu.edu; ramin.tafazzoli@tamu.edu; ehsani@ece.tamu.edu

*Abstract*— DC-DC converters with galvanic isolation are a key element in the aircraft DC distribution system and Dual and Triple Active Bridge converters are one of the most interesting candidates in this application. The high-frequency transformer leakage inductances play a key role in the AC Link of these converters. This leakage inductances determine the power transfer capability of the converter and shape the AC link currents. The leakage inductance value is related to the distribution of the transformer windings and it changes with the RMS value of the AC link current. In this paper, first a high frequency transformer is designed for a Dual and Triple Active Bridge converter for the More Electric Aircraft DC power system. Then, an Ansys/Maxwell Finite Element analysis is performed on the leakage inductances of the transformer in three different winding configurations and in different AC link RMS current values. Finally, the transient performance of the design is validated by the Ansys/Maxwell Transient.

*Keywords*— *Dual Active Bridge, Triple Active Bridge, High Frequency Transformer, Multi-port Transformer, AC Link, Leakage Inductance, More Electric Aircraft*

## I. INTRODUCTION

Electrification of the aircraft is helpful in reducing mass, fuel consumption and environmental impacts, as well as in power and energy management. Furthermore, electrification can increase reliability and the flexibility in power generation and distribution [1]. DC power distribution system is one the most promising architectures in the More Electric Aircraft (MEA), because of the possibility of making easier couplings between the main source and the storage devices and its lower overall mass in comparison to AC distribution architectures [2]. High power DC-DC converters with galvanic isolation are a key element in the aircraft DC distribution system, electric vehicle power system, charging stations, etc [2-5]. Dual Active Bridge (DAB) converters are one of the most interesting candidates in this application due to their capability to work with Zero Voltage Switching (ZVS) over a wide operating range and without a significant increase in the control complexity [6]. As can be seen from Fig.1, the high frequency transformer in the AC link of the DAB converter is responsible to make the galvanic isolation and to transfer power from the primary side of the converter to the secondary. According to (1), the inductance of the AC link is one of the parameters which determines the amount of the transferred power. The smaller the inductance of the AC link, the higher the power transfer capability of the converter. However, the amount of the AC link inductance cannot be very small as the current of the converter shapes by the AC link inductance. It can help to achieve ZVS, reduce the peak value of the AC link current and subsequently increase the efficiency of the converter [7]. The amount of the AC link inductance is related to the leakage inductance of the high frequency transformer and if the leakage inductance of the transformer is very small, an extra inductor will be needed for shaping the current and to reach ZVS. Therefore, the amount of the leakage inductance of the high frequency transformer plays an important role in this converter and a proper design of the transformer can avoid using an extra inductor in the AC link of the converter. In [7-11], a multi-port high frequency transformer is designed for the solid-state transformer application and, in [12-14], a high frequency transformer is designed for the Dual Active Bridge converter. However, the leakage inductance of the transformer is not considered as a main design objective in their design and except some comparisons in some papers there is not a comprehensive analysis on the value of the leakage inductance in these studies. In [15-16], an analysis is performed on the transformer leakage inductance of Triple and Dual Active Bridges. However, the effect of the AC link RMS current is not considered in these studies.

In this paper, first a high frequency transformer is designed for a Dual and Triple Active Bridge according to the MEA DC power system characteristics. Then, a magnetostatic Ansys/Maxwell Finite Element analysis is performed on the leakage inductances of the transformer in three different winding configurations and in different AC link RMS current values. Finally, the transient performance of the design is validated by the Ansys/Maxwell Transient.

## II. DUAL ACTIVE BRIDGE PRINCIPAL AND TRANSFORMER DESIGN

### A. Dual Active Bridge Principal

Fig. 1 shows a schematic of the Dual Active Bridge converter circuit. It includes two full-bridge inverters which are connected through a high frequency transformer in the AC link. Conventionally, each one of the full bridges produces a square-wave voltage and, according to (1), the power flow between the bridges is determined by the primary and secondary square-wave voltage amplitudes $V_1$ and $V_2$, the switching frequency of the converter $f_s$, the AC link inductance $L$, and the phase shift between the primary and secondary square-wave voltages $\varphi$ [17]. In the conventional control methods of this converter, the only parameter that controls the power flow is the phase shift. However, the switching frequency and the AC link inductance are two important parameters as well that need to be determined in the design procedure of the converter.

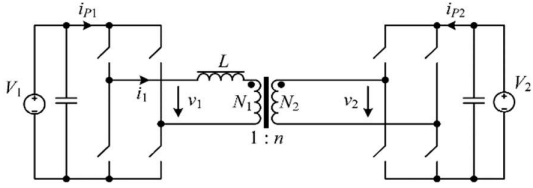

Fig. 1. Dual Active Bridge Converter

$$P_{12} = \frac{nV_1V_2}{2\pi f_s L}\varphi\left(1 - \frac{|\varphi|}{\pi}\right) \quad (1)$$

The maximum power is transferred when $\varphi = \frac{\pi}{2}$

*B. Dual Active Bridge Transformer Design*

There is an approach called the core geometrical constant $K_{gfe}$ approach for designing transformers which is used when the maximum flux density of the core is to be chosen to minimize the total power loss of the transformer. Based on this approach, first the core geometrical constant must be calculated using (2), then according to the calculated $K_{gfe}$, a proper core must be selected and the maximum flux density $B_{max}$ must be evaluated using (3). Finally, the primary and secondary number of turns $N_1$ and $N_2$, the fraction of winding area $\alpha_1$ and $\alpha_2$ for each winding and the windings wire sizes $A_{w1}$ and $A_{w2}$ must be calculated using (4)-(6) [18].

$$K_{gfe} \geq \frac{\rho \lambda^2 I_{rmsT}^2 K_{fe}^{(2/\beta)}}{4 K_u P_{loss}^{(\beta+2/\beta)}} 10^8 \quad (2)$$

$$B_{max} = \left[\frac{\rho \lambda^2 I_{rmsT}^2}{2K_u}\frac{(MLT)}{W_A A_c^3 l_m}\frac{1}{\beta K_{fe}}\right]^{(1/\beta+2)} \quad (3)$$

$$N_1 = \frac{\lambda}{2B_{max}A_c}10^4, \; N_2 = nN_1 \quad (4)$$

$$\alpha_1 = \frac{N_1 I_{rms1}}{N_1 I_{rmsT}}, \alpha_2 = \frac{N_2 I_{rms2}}{N_1 I_{rmsT}} \quad (5)$$

$$A_{w1} \leq \frac{\alpha_1 K_u W_A}{N_1}, A_{w2} \leq \frac{\alpha_2 K_u W_A}{N_2} \quad (6)$$

In these equations, $\lambda$ is the volt-seconds applied to the primary winding during the positive portion, $K_{fe}$ is the core loss coefficient, $P_{loss}$ is the transformer total loss, $A_c$ is the core cross section area, $\beta$ is the core loss exponent, $l_m$ is the magnetic length, MLT is Mean Length per Turn, $W_A$ is the core winding area, n is the transformer turns ratio, $I_{rms1}$ is the primary rms current, $I_{rms2}$ is the secondary rms current, $I_{rmsT}$ is the total rms current referred to the primary side, $K_u$ is the winding fill factor and $\rho$ is the wire resistivity.

According to the above procedure, a transformer is designed for a DAB converter to supply a load at the 27 VDC bus of the MEA power system from its 270 VDC bus. The design parameters are summarized in Table.1.

Table.1 Dual Active Bridge Transformer Design Parameters

| Parameter | Value |
|---|---|
| Power Rating | 1 kW |
| Primary Voltage Amplitude | 270 V |
| Secondary Voltage Amplitude | 27 V |
| Switching Frequency | 20 kHz |
| Duty Cycle | 50% |
| Total AC Link Leakage Inductance | 0.227 mH |
| Turns Ratio | 10 |
| Allowed Total Power Dissipation | 2% |

As the result, the PC47EE57/47-Z EE core from TDK company is selected as the core of the transformer and the results of the design with this core at the phase shift of $\varphi = 0.4581 \; rad$ are summarized in Table. 2.

Table. 2 Dual Active Bridge Transformer Specifications

| Parameter | Value |
|---|---|
| Core Material | Ferrite |
| Winding Fill Factor | 50% |
| Core Maximum Flux Density | 0.1962 T |
| Primary Number of Turns | 50 |
| Secondary Number of Turns | 5 |
| Primary Fraction of Winding Area | 50% |
| Secondary Fraction of Winding Area | 50% |
| Primary Winding AWG# | 16 |
| Secondary Winding AWG# | 6 |

As can be seen from Table. 2, the maximum flux density in the core, 0.1962 T, is less than the saturation limit of the ferrite which is 0.35 T. So, the transformer works in the linear region of the core.

According to Table. 1, a 0.227 mH inductance is required as the total inductance of the converter AC link. This inductance value can be achieved either using only the leakage inductances of the transformer or an extra inductor may be needed in the AC link. To find a proper option, first the leakage inductances of the transformer must be calculated. According to (7), the leakage inductances of a transformer can be found using the self and mutual inductances of the transformer [18].

$$L_{l1} = L_{11} - \left(\frac{N_1}{N_2}\right)L_{12} \quad (7)$$

$$L_{l2} = L_{22} - \left(\frac{N_2}{N_1}\right)L_{21}$$

$$L_{lT} = L_{l1} + L_{l2}\left(\frac{N_1}{N_2}\right)^2$$

In (7), $L_{11}$ and $L_{22}$ are the self-inductances of the primary and secondary windings, $L_{12}$ and $L_{21}$ are the mutual inductances between them, $L_{l1}$ and $L_{l2}$ are the primary and secondary leakage inductances and $L_{lT}$ is the total leakage inductance of the transformer referred to the primary side.

To avoid using an extra inductor in the AC link, the required inductance of the converter AC link needs to be provided by only the total leakage inductance of the transformer. One degree of freedom to increase or decrease the leakage inductances of the transformer is playing with the mutual inductance value, as the self-inductances are fixed by the number of turns and the geometry of the core. The value of the leakage inductance can be increased by adding an air gap in the core as well. However, there will be a loss in the flux of the core in this way. Therefore, to change the mutual inductance value, a better way is to change the displacement of the windings respect to each other. The effect of the windings' displacement on the leakage inductance of the transformer are studied in the section IV.

## III. TRIPLE ACTIVE BRIDGE PRINCIPAL AND TRANSFORMER DESIGN

### A. Triple Active Bridge Principal

The Dual Active Bridge converter can be extended to a Triple Active Bridge (TAB) by adding another full bridge as the third port and another winding to the high frequency transformer which has couplings with the other windings as it is shown in Fig. 2.

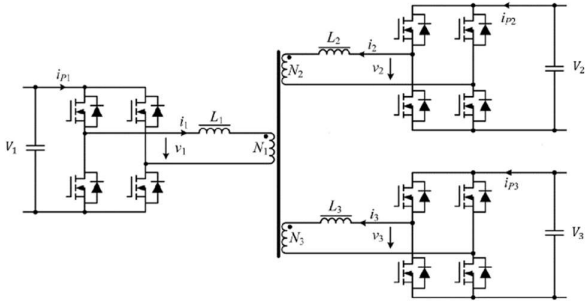

Fig. 2. Triple-Active-Bridge Converter

In order to control and analysis of this converter, it needs to be modeled as a delta connection as it is shown in fig. 3 [19].

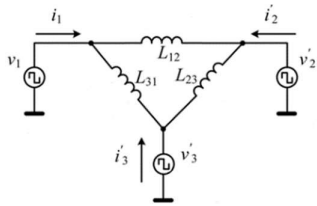

Fig. 3. Triple-Active-Bridge Converter Delta Equivalent circuit

Fig. 3 is the delta equivalent circuit of the converter referred to port 1. As can be seen from the figure, each port is connected to the others through an inductance. Therefore, according to (1), the amount of the power transferred between each two ports can be determined by the inductance value and the phase shift between those ports. The amount of $L_{12}$, $L_{13}$ and $L_{23}$ can be calculated by (8).

$$L_{12} = \frac{L_{l1}L'_{l2} + L_{l1}L'_{l3} + L'_{l2}L'_{l3}}{L'_{l3}} \quad (8)$$

$$L_{13} = \frac{L_{l1}L'_{l2} + L_{l1}L'_{l3} + L'_{l2}L'_{l3}}{L'_{l2}}$$

$$L_{23} = \frac{L_{l1}L'_{l2} + L_{l1}L'_{l3} + L'_{l2}L'_{l3}}{L_{l1}}$$

Where $L_{l1}$, $L'_{l2}$ and $L'_{l3}$ are the leakage inductance of each port of the transformer in its star equivalent circuit referred to port 1 [19] and they can be calculated by (9) using the self and mutual inductances of the transformer.

$$L_{l1} = L_{11} - (\frac{N_1}{N_2})L_{12} \quad (9)$$
$$L_{l2} = L_{22} - (\frac{N_2}{N_1})L_{21}$$
$$L_{l3} = L_{33} - (\frac{N_3}{N_1})L_{13}$$

### B. Triple Active Bridge Transformer Design

Using the same design procedure and approach in section II, a transformer is designed for a TAB connecting the MEA 270 VDC bus, 27 VDC bus and a battery pack of 135 V. The design parameters are mentioned in Table. 3, and the resulted specifications of the designed transformer are summarized in Table. 4.

Table.3 Triple Active Bridge Transformer Design Parameters

| Parameter | Value |
|---|---|
| Power Rating | 1 kW |
| Port 1 Voltage Amplitude | 270 V |
| Port 2 Voltage Amplitude | 27 V |
| Port 3 Voltage Amplitude | 135 V |
| Switching Frequency | 20 kHz |
| Duty Cycle | 50% |
| $L_{12}=L_{13}=L_{23}$ | 0.227 mH |
| Allowed total power dissipation | 2% |

Table. 4 Triple Active Bridge Transformer Specifications

| Parameter | Value |
|---|---|
| Core Type | TDK PC47EE57/47-Z EE core |
| Core Material | Ferrite |
| Winding Fill Factor | 75% |
| Core Maximum Flux Density | 0.1962 T |
| Port 1 Winding Number of Turns | 50 |
| Port 2 Winding Number of Turns | 5 |
| Port 3 Winding Number of Turns | 25 |
| Port 1 Winding Fraction of Winding Area | 33% |
| Port 2 Winding Fraction of Winding Area | 33% |
| Port 3 Winding Fraction of Winding Area | 33% |
| Port 1 Winding AWG# | 16 |
| Port 2 Winding AWG# | 6 |
| Port 3 Winding AWG# | 13 |

As can be seen from Table. 3, it is desirable to have equal inductances in the delta equivalent circuit of the converter. Having the same inductances results to have equal transferred powers with equal phase shifts between each two ports of the converter. Otherwise, in order to have balanced loads on the ports, different phase shift values must be applied to the ports which leads to have circulating powers and losses. It will be examined in the next section whether equal inductances can be achieved using EE cores or not.

## IV. FINITE ELEMENT SIMULATION AND RESULTS

Three different configurations of the transformer with different windings' displacement are shown in Fig. 4. These configurations are modeled in Ansys/Maxwell and as the result of the magnetostatic analysis of the software, the variation of the leakage inductance with the RMS current in the primary side is evaluated for each configuration and shown in Fig. 5.

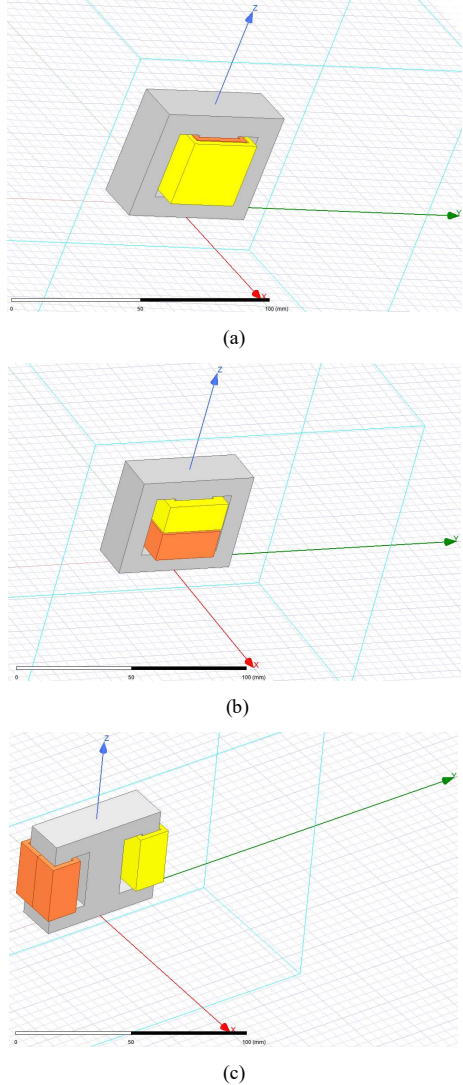

Fig. 4. Different Winding Distributions in the Core

In the concentric configuration shown in Fig. 4(a), the windings are very close to each other and there is a strong coupling between their fields. Therefore, it is expected to have a mutual inductance comparable to the self-inductances and as the result a small leakage inductance in this configuration. Fig. 5(a) validates a small leakage inductance in this configuration, around 10 times smaller than the design requirement, in every RMS current value. In Fig. 4(b), windings are placed in a row in the middle branch of the core. In this configuration, the distance between the windings is more than the concentric configuration which results a weaker coupling, a smaller mutual inductance, and a higher leakage inductance. This is validated in Fig. 5(b). As can be seen from this figure, the amount of the leakage inductance is higher than the concentric configuration and it is very close to the design requirement in all RMS currents. In the last configuration in Fig. 4(c), the windings are placed in the farthest distance respect to each other which results a very weak coupling, a very small mutual inductance and a very high leakage inductance as can be seen from Fig. 5(c).

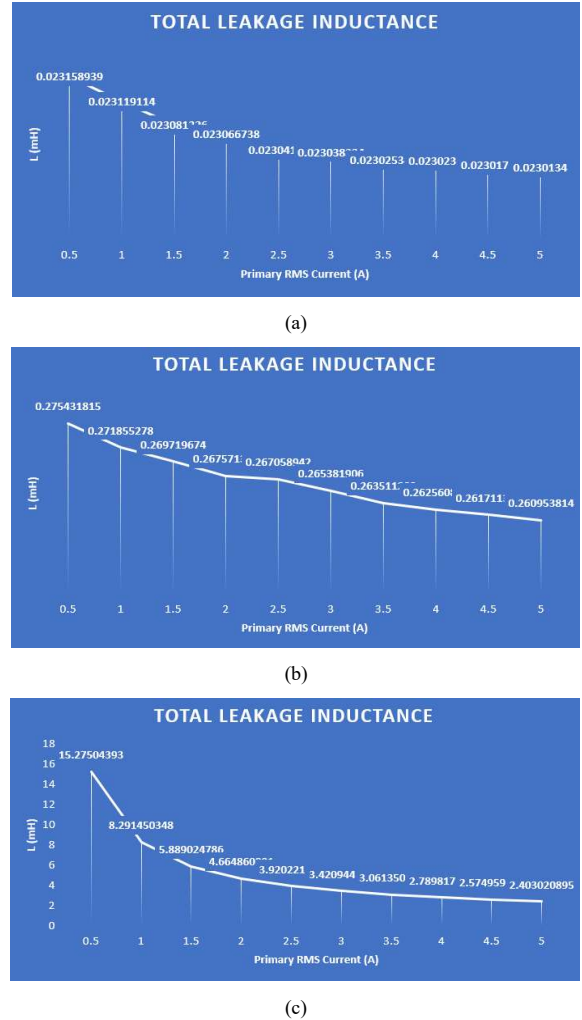

Fig. 5. Variation of Leakage Inductance with RMS Current

One more thing that can be found from Fig. 5 is that in the concentric configuration and in the configuration of Fig. 4(b), the variation of the leakage inductance with the RMS current is too small and the amount of the leakage inductance can be considered as a fixed value in all RMS currents. However, this is not valid for the configuration of Fig. 4(c). As can be seen from this figure, there is a big change in the amount of the leakage inductance with changing the RMS current specially in low RMS currents which is problematic for controlling the power flow in the converter.

All these discussions can be verified by the magnetizing inductance curves in Fig. 6 as well. As can be seen from the figure, the amount of the magnetizing inductance in configurations (a) and (b) is much higher than the configuration (c) which shows a stronger coupling, a higher mutual inductance and consequently a lower leakage inductance in (a) and (b) configurations.

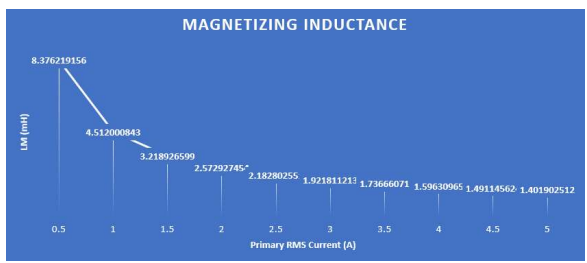

(a)

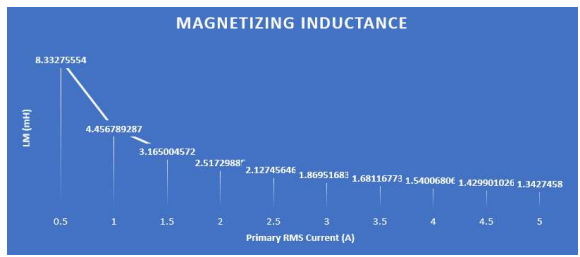

(b)

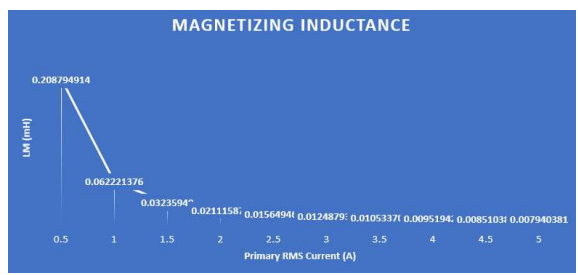

(c)

Fig. 6. Variation of Magnetizing Inductance with RMS Current

As the summary of this section, although the leakage inductance value is robust against the variation of the RMS current in both Fig. 4(a) and Fig. 4(b) configurations, the amount of the leakage inductance in the concentric configuration is too small to shape the current in the AC link. The amount of the leakage inductance in the configuration of Fig. 4(c) is too high which leads to a low power transfer capability for the converter. In addition, the variations of the leakage inductance with the RMS current in this configuration results a serious problem in controlling the power flow in the converter. As the result, since the average value of the leakage inductance in the Fig. 4(b) configuration (0.26 mH) is very close to the design requirement and since the variation of the leakage inductance with the RMS current in this configuration is too small, this configuration seems to be the best candidate for the design requirement.

For the validation of the design, a transient analysis is performed by the Ansys/Maxwell Transient and the result is shown in Fig. 7.

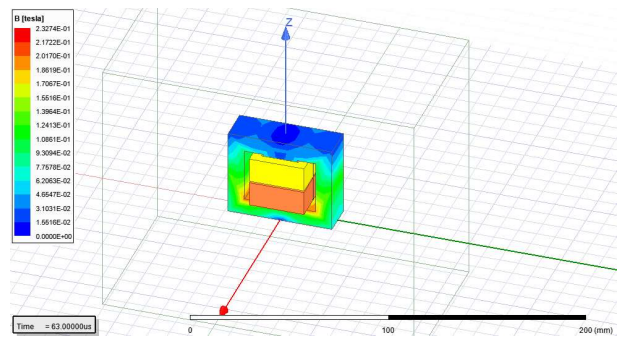

Fig. 7. Flux Density Distribution in the Core

As can be seen from the figure, the maximum flux density in the core is 0.23274 T which is less than the saturation limit of the ferrite and it is close to the designed value.

Now that the configuration of Fig. 4(b) is selected as the best candidate for DAB for the design requirement, the designed three windings transformer for TAB can be simulated with this winding configuration to examine if equal inductances can be achieved in the delta equivalent circuit using EE core. Fig. 8 shows the simulated model of the transformer with three windings with the specifications mentioned in Table. 4. The result of the simulation for 5A primary RMS current and for the case that the winding 2 is between windings 1 and 3 is shown in Table. 5.

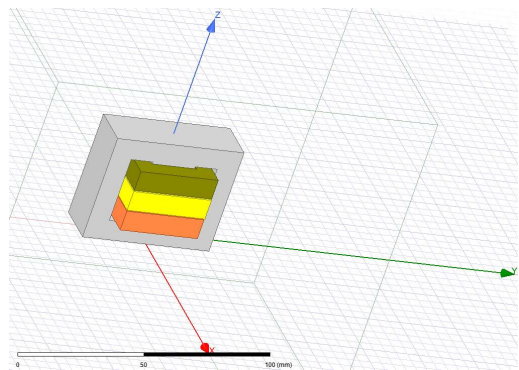

Fig. 7. Three Winding Transformer Configuration

Table. 5 TAB Delta Equivalent Circuit Inductances

| $L_{12}$ (mH) | $L_{13}$ (mH) | $L_{23}$ (mH) |
| --- | --- | --- |
| 0.208612 | 0.576136 | 0.39146 |

As can be seen from Table. 5, the amount of the inductances in the TAB delta equivalent circuit are not the same using this core and winding configuration. Because, according to the result of the simulation for DAB, $L_{13}$ has a higher inductance value than the others since winding 2 is in between windings 1 & 3 and the distance between windings 1 and 3 is more than the others. As the result, the EE core is not a good candidate for TAB application and a core with symmetric winding configuration like toroidal core is required for this application.

## V. Conclusions

In this paper, first, a high frequency transformer is designed for a Dual and Triple Active bridge converter according to the More Electric Aircraft DC power system parameters. Then, an analysis is performed on the leakage inductance values of the transformers and their relationship with the core shape, winding distribution, and the converter AC link RMS current. Finally, an Ansys/Maxwell simulation is performed for the validation of the analysis. As the result, for the Dual Active Bridge, in the case that the windings are far away from each other, the value of the leakage inductance of the transformer is much higher than the case they are close. In addition, the variation of the leakage inductance value with the AC link RMS current is much higher when windings are far apart. For the Triple Active Bridge, it is concluded that the EE cores are not an appropriate option to achieve equal inductances in the delta equivalent circuit of the converter and a core with a symmetric winding configuration like toroidal cores is preferred for this application.